\documentclass[showpacs,aps,graphicx,twocolumn]{revtex4}%
\usepackage{graphicx}

\begin{document}
\title{Multiparty quantum secret splitting  and quantum state sharing}
\author{ Fu-Guo Deng,$^{1,2,3}$ Xi-Han Li,$^{1,2}$ Chun-Yan Li,$^{1,2}$ Ping Zhou,$^{1,2}$
and Hong-Yu Zhou$^{1,2,3}$}
\address{$^1$ The Key Laboratory of Beam Technology and Material
Modification of Ministry of Education, Beijing Normal University,
Beijing 100875,
People's Republic of China\\
$^2$ Institute of Low Energy Nuclear Physics, and Department of
Material Science and Engineering, Beijing Normal University,
Beijing 100875, People's Republic of China\\
$^3$ Beijing Radiation Center, Beijing 100875,  People's Republic
of China}
\date{\today }

\begin{abstract}

A protocol for multiparty quantum secret splitting is proposed
with an ordered $N$ Einstein-Podolsky-Rosen (EPR) pairs and
Bell-state measurements. It is secure and has the high intrinsic
efficiency and source capacity as almost all the instances are
useful and each EPR pair carries two bits of message securely.
Moreover, we modify it for multiparty quantum state sharing of an
arbitrary $m$-particle entangled state based on quantum
teleportation with only Bell-state measurements and local unitary
operations which make this protocol more convenient in a practical
application than others.

\end{abstract}
\pacs{03.67.Hk, 03.67.Dd, 03.65.Ud, 89.70.+c} \maketitle

\section{introduction}

Quantum information, an ingenious application of quantum mechanics
within the field of information has attracted a lot of attentions
\cite{book,longcomputation,teleportation,TE,BW}. In particular
almost all the branches of quantum communication have been
developed quickly since the original protocol was proposed by
Bennett and Brassard \cite{BB84} in 1984, such as quantum key
distribution (QKD)
\cite{BB84,Gisinqkd,Cabello,longqkd,dengqkd,Ekert91,BBM92,Hwang,ABC},
quantum secure direction communication (QSDC)
\cite{two-step,QOTP,Wangc}, quantum teleportation
\cite{teleportation,TE}, quantum secret sharing (QSS)
\cite{HBB99,KKI,Bandyopadhyay,Karimipour,deng2005,deng20052,longqss,
guoqss,zhanglm,improving,zhangPLA,YanGao,cleve,Peng,dengmQSTS,dengpra,Gottesman,TZG,AMLance},
and so on. QKD provides a secure way for creating a private key
between two remote parties. With a private key, most of the goals
in classical secure communication can be accomplished. For
example, a classical secret message can be transmitted securely by
using classical one-time pad crypto-system with a private key. To
date, QKD has progressed quickly and becomes one of the most
mature applications of quantum information.

QSS is likely to play a key role in both transmitting of a
classical information and protecting a secret quantum information,
such as in secure operations of distributed quantum computation,
sharing difficult-to-construct ancillary states and joint sharing
of quantum money \cite{Gottesman,zhangPLA}. Suppose a banker,
Alice, wants her two agents, Bob and Charlie who are at remote
places to deal with her business according to her message ($M_A$).
However Alice doubts that one of them may be dishonest and the
business may be destroyed if the dishonest one manages it
independently. Moreover Alice does not know who the dishonest one
is, but she knows that the number of dishonest person is less than
two. To prevent the dishonest man from destroying the business,
classical cryptography provides the secret sharing scheme
\cite{Blakley} in which Alice splits her message ($M_A$) into two
sequences ($M_B$ and $M_C$) and sends them to Bob and Charlie,
respectively. They can read out the message $M_A=M_B\oplus M_C$ if
and only if they cooperate, otherwise they will get nothing.

QSS is the generalization of classical secret sharing
\cite{Blakley} into quantum scenario and supplies a secure way for
sharing not only a classical information
\cite{HBB99,KKI,guoqss,longqss,deng2005,deng20052}  but also a
quantum information
\cite{cleve,Bandyopadhyay,Peng,Karimipour,zhanglm,dengmQSTS,dengpra,Gottesman}.
In the latter, the sender, Alice will send an unknown quantum
state to her $m$ agents and one of them can recover it with the
help of the others \cite{dengmQSTS}. Compared with QKD, QSS can
reduce the the resources necessary for the communication
\cite{guoqss} for a classical secret information. Moreover, the
sharing and the splitting of an unknown quantum state  should
resort to QSS with quantum entanglement, which is called quantum
state sharing \cite{AMLance} (abbreviate it as QSTS for the
difference from QSS of a classical information \cite{dengmQSTS})
recently, and cannot be implemented in only  a classical way. For
example, Li et al. proposed a QSTS scheme for sharing an unknown
single qubit with $m$ agents. In their scheme, Alice shares $m$
Einstein-Podolsky-Rosen (EPR) pairs with the $m$ agents, and she
performs an $(m+1)$-particle Greenberger-Horne-Zeilinger (GHZ)
state measurement on the unknown quantum system and her $m$ EPR
particles. $m-1$ agents take a single-particle measurement on
their EPR particles and the last agent can recover the original
unknown state with the helps of all the other agents.

Quantum teleportation \cite{teleportation,TE} is a quantum
technique with which the two remote parties, the sender Alice and
the receiver Bob, exploit the nonlocal correlations of the quantum
channel shared in advance, such as an Einstin-Podolsky-Rosen (EPR)
pair, to teleport an unknown quantum state $\vert
\phi\rangle_u=\alpha\vert \uparrow\rangle + \beta\vert \downarrow
\rangle$. The task can be accomplished by means that Alice makes a
Bell-basis measurement on her EPR particle and the unknown quantum
system $u$, and Bob reconstructs the state $\vert \phi\rangle_u$
with a local unitary operation on his EPR particle according to
the classical information published by Alice
\cite{teleportation,TE}.

In this paper, we will propose a protocol for multiparty quantum
secret splitting with EPR pairs following some ideas in Refs.
\cite{two-step,Wangc,zhangPLA}. It has the high intrinsic
efficiency and source capacity as almost all the instances are
useful and each EPR pair carries two bits of message securely.
Moreover, we modify it for multiparty quantum state sharing
(MQSTS) of an arbitrary $m$-particle entangled state based on
quantum teleportation \cite{Yangcp}. The parties exploit EPR pairs
and Bell-state measurements to accomplish this task, which makes
this protocol more convenient in a practical application than
others.

\section{multiparty quantum secret splitting  with  EPR pairs}
In a QSS, its security is simplified to prevent the dishonest
agent from eavesdropping freely as the parties can detected any
eavesdropper if they can find out the dishonest one in the agents
\cite{KKI}. Our multiparty quantum secret splitting (MQSSP)
protocol inherits this feature. For convenience, let us first
describe a three-party quantum secret splitting protocol, and then
generalize it to the case with $N$ agents, similar to Ref.
\cite{zhangPLA}.

An EPR pair is in one of the four Bell states shown as following
\cite{book}:
\begin{eqnarray}
\left\vert \psi ^{\pm}\right\rangle_{AB}
=\frac{1}{\sqrt{2}}(\left\vert 0\right\rangle _{A}\left\vert
1\right\rangle _{B}\pm\left\vert
1\right\rangle _{A}\left\vert 0\right\rangle _{B}), \label{EPR12}\\
\left\vert \phi ^{\pm}\right\rangle_{AB}
=\frac{1}{\sqrt{2}}(\left\vert 0\right\rangle _{A}\left\vert
0\right\rangle _{B}\pm\left\vert 1\right\rangle _{A}\left\vert
1\right\rangle _{B}). \label{EPR34}
\end{eqnarray}
where $\vert 0\rangle$ and $\vert 1\rangle$ are the two
eigenvectors of a two-level quantum system, and the subscripts $A$
and $B$ represent the two particles in an EPR pair. The four local
unitary operations $U_{i}$ ($i=0,1,2,3$) can transform one of the
Bell states into another.
\begin{eqnarray}
U_{0}&&\equiv I=\vert 0\rangle \langle 0\vert + \vert 1\rangle
\langle 1\vert , \label{O0}
\\
U_{1}&&\equiv \sigma _{z}=\vert 0\rangle \langle 0\vert -\vert
1\rangle \langle 1\vert , \label{O1}
\\
U_{2}&&\equiv \sigma _{x}=\vert 0\rangle \langle 1\vert + \vert
1\rangle \langle 0\vert , \label{O2}
\\
U_{3}&&\equiv i\sigma _{y}=\vert 0\rangle \langle 1\vert -\vert
1\rangle \langle 0\vert, \label{O3}
\end{eqnarray}
where $I$ is the $2\times 2$ identity matrix and $\sigma_i$ are
the Pauli matrices.

Suppose the three parties in the MQSTS are the sender Alice and
her two agents, Bob and Charlie. The basic idea of our MQSTS
protocol is that the two photons in each EPR pairs prepared by the
agent Bob are not transmitted simultaneously in the insecure
quantum channel, and Alice codes her message with four local
unitary operations after confirming the security of the channel,
same as the two-step QSDC scheme \cite{two-step}. We can describe
our MQSTS protocol in detail as follows.

(1) Alice, Bob and Charlie agree on that the four local unitary
operations $U_0$, $U_1$, $U_2$ and $U_3$ represent the two bits of
information 00, 11, 01 and 10, respectively. They use the two
measuring bases (MBs), $Z\equiv\{$$\vert 0\rangle$, $\vert
1\rangle\}$ and $X\equiv\{$$\vert
+x\rangle=\frac{1}{\sqrt{2}}(\vert 0\rangle + \vert 1\rangle)$,
$\vert -x\rangle=\frac{1}{\sqrt{2}}(\vert 0\rangle - \vert
1\rangle)\}$ (for instance, the polarizations of a photon along
$z-$ and $x-$ directions), to measure the sample photons randomly
in the process of eavesdropping check.

(2) The agent Bob prepares an ordered $N$ EPR polarization photon
pairs in the same quantum state, say $ \vert \psi^- \rangle _{AC}
=\frac{ 1}{\sqrt{2}}(\vert 0\rangle _{A}\vert 1\rangle _{C} -
\vert 1\rangle _{A}\vert 0\rangle _{C})$. We
denote the N ordered EPR pairs with [(P$_{1}$($A$),P$_{1}$($C$)), (P$_{2}$($%
A $),P$_{2}$($C$)), (P$_{3}$($A$),P$_{3}$($C$)), ..., (P$_{N}$($A$),P$_{N}$(%
$C$))], same as Refs. \cite{longqkd,two-step,QSDCNetwork}.

(3) Bob takes one photon from each EPR pair to form an ordered EPR
partner photon sequence, say [P$_{1}$($A$), P$_{2}$($A$),
P$_{3}$($A$), ... , P$_{N}$($A$)], called the $S_A$ sequence. The
remaining EPR partner photons compose another EPR partner
photon sequence [P$_{1}$($C$), P$_{2}$($C$), P$_{3}$($C$), ... , P$_{N}$($%
C$)], which is called the $S_C$ sequence.

(4) Bob fist sends the $S_A$ sequence to Alice and keeps the $S_C$
sequence. Alice picks out a sufficiently large subset of photons
from the sequence $S_A$ for the eavesdropping check of the
transmission.

The check can be completed with the following procedures: (a)
Alice tell Bob which photons he has chosen and Bob picks out the
correlated photons in the sequence $S_C$. (b) Bob chooses randomly
the measuring basis (MB) $Z$ or $X$ to measure the chosen photons.
(c) Bob tells Alice which MB he has chosen for each photon and the
outcomes of his measurements. (d) Alice uses the same MBs as Bob
to measure the corresponding photons and checks the eavesdropping
with the results of Bob's. If no eavesdropping exists, their
results should be completely opposite in an ideal condition, i.e.,
if Alice gets 0 (1), then Bob gets 1 (0). This is the first
eavesdropping check.

After that, if the error rate $\varepsilon_1$ is small, Alice and
Bob can conclude that there are no eavesdroppers in the line.
Alice and Bob continue to perform step 5; otherwise they have to
discard their transmission and abort the communication.

(5) Bob chooses randomly one of the four local unitary operations
$U_i$ ($i=0,1,2,3$) to encrypt each of the photons in the sequence
$S_C$, say $U_B$, and then he sends the sequence $S_C$ to Alice.

(6) Alice analyzes the error rate $\varepsilon_2$ of the
transmission of $S_C$. This is the second eavesdropping check. For
this end, Alice first picks up $k+j$ EPR photon pairs and then
requires Bob to publish the operations $U_B$ with which he has
operated on these pairs. Alice takes the single-photon
measurements on the two photons of $k$ pairs by choosing randomly
the MB X or Z. For the other $j$ pairs, she only measures one
photon in each pair with the MB Z or X chosen randomly. She uses
the other photons in these pairs as the decoy photons for checking
eavesdropping in the next step.

(7) Alice inserts randomly the decoy photons in the sequence $S_C$
and then sends it to Charlie. They analyze the security of this
transmission with the decoy photons. If they conclude that the
quantum line is secure, they continue their communication to next
step; otherwise, they discard the results and repeat the quantum
communication from the beginning.

(8) Alice selects a subset of photons as the samples for
eavesdropping check and chooses one of the four unitary operations
randomly on each sample. For other photons in the $S_A$ sequence
(except for those for eavesdropping check), Alice encodes her
message $M_A$ on them with the four unitary operations, and then
sends the sequence $S_A$ to Charlie.

(9) Charlie performs the Bell measurements on the EPR photon pairs
and reads  out the combination of the operations done by Alice and
Bob, i.e., $U_C = U_A \otimes U_B$.

(10) Alice and Charlie finish the error rate of the samples
selected by Alice, and determine whether there are eavesdroppers
monitoring the quantum channel when the $S_A$ is transmitted from
Alice to Charlie. For this end, Alice first requires Bob to
publish the operations done on the correlated photons in the
sequence $S_C$, and then she requires Charlie to tell her the
results of the Bell measurements. If the transmission of the $S_A$
sequence is secure, Alice tells Bob and Charlie to collaborate for
reading out the message $M_A$, otherwise they will abandon the
results of the transmission.

This MQSSP protocol follows some ideas in the two-step QSDC scheme
\cite{two-step} and the Bennett-Brassard-Mermin 1992 (BBM92) QKD
protocol \cite{BBM92}. The process for the transmission of the
$S_A$ sequence from Bob to Alice is same as that in Ref.
\cite{two-step}, and its security is same as the BBM92 QKD
protocol \cite{BBM92} whose security is proven in both  an ideal
condition \cite{IRV} and a practical condition \cite{WZY}. So the
process for the transmission of the $S_A$ sequence is secure for
any eavesdropper including the agent Charlie. The operation done
by Bob on each photon in the $S_C$ is equivalent to the encryption
on the photon with a random key, which makes any other one have no
ability for reading out the information on the photon, same as the
quantum one time pad \cite{Gisinqkd,QOTP}. That is, any
eavesdropper except the agent Bob cannot eavesdrop the message
$M_A$ even though he monitors the quantum channel in the
subsequent processes.

The goal that the $S_C$ sequence is first transmitted to Alice
before Alice encodes her message on the $S_A$ sequence is to
prevent Bob from eavesdropping with a fake signal and cheat
freely. The eavesdropping check of the transmission of the  $S_C$
sequence is necessary for Alice and Charlie with some decoy
photons. Also, the cheating of Bob's will be found out by
comparing the results of the measurements on the decoy photons.
After the processes for the secure transmission of the $S_A$
sequence from Bob to Alice and that of the $S_C$ sequence from
Alice to Charlie, any one cannot steal the message $M_A$ except
that Bob and Charlie cooperate. Then the present MQSSP protocol is
secure.

In fact, the security in each procedure of the transmission of the
particle sequences is ensured by the eavesdropping check in this
scheme, which is different to Ref. \cite{zhangPLA}. In the latter,
the agent Bob does not sample his particles and measure them,
which makes it insecure if the other agent Charlie wants to steal
Bob's information with multi-photon fake signal, similar to Ref.
\cite{improving}. That is, Charlie intercepts the signal
transmitted from Alice to Bob, and sends  some photons in the EPR
pairs with same Bell state in a time slot. Charlie can  obtain
almost all the information about the operations done by Bob with
photon number splitters. Moreover, she can read out the
information with a suitable delay Trojan horse attack
\cite{dengattack}. This case does not happen in this scheme.

The generalization of this MQSSP protocol to the case with $N$
agents can be implemented in a simple way by modifying the
processes in the case with two agents. We describe it after the
step 4 discussed above.

(5') Bob chooses randomly one of the four local unitary operations
$U_i$ ($i=0,1,2,3$) to encrypt each of the photons in the sequence
$S_C$, say $U_B$, and then he sends the sequence $S_C$ to  next
agent, Charile.

(6') Alice and Charlie analyze the error rate of this transmission
by choosing randomly a subset of the EPR pairs and requires Bob to
tell them his operations $U_B$ on these samples. They measure
their sample photons with the correlated MBs, same as that between
Alice and Bob. If the error rate is lower than the threshold
value, Charlie operates the photons in the sequence $S_C$ by
choosing randomly one of the four operations $U_i$ ($i=0,1,2,3$).
Also, he picks up some samples from  $S_C$ and performs a Hadamard
operation on each sample. He sends the sequence to the next agent
Dick.

(7') Alice and Dick complete the eavesdropping check of this
transmission, same as that between Alice and Charlie. The
difference is that Charlie should attend to check eavesdropping in
this time. He publishes some positions of the the photons operated
with Hadamard operations, and then Alice and Dick confirm whether
there is an eavesdropper who has stolen the information about the
operations done by Charlie. If the quantum communication is still
secure, Dick repeats the procedure of Charlie's and sends the
sequence $S_C$ to the next agent.

(8') After repeating the step 7' N-3 times, the $S_C$ sequence is
received securely by Yang, the agent before the last one Zach.
After the operations done by Yang, similar to Charlie, he sends
the sequence $S_C$ to Alice (Not Zach!).

(9') Alice analyzes the error rate of the transmission of $S_C$
between her and Yang. She first picks up $k+j$ EPR photon pairs
and then requires all the agents except for the last one to
publish their operations with which he has operated on these
pairs. Alice takes the single-photon measurements on the two
photons of $k$ pairs by choosing randomly the MB X or Z. For the
other $j$ pairs, she only measures one photon in each pair with
the MB Z or X chosen randomly. She uses the other photons in these
pairs as the decoy photons for checking eavesdropping in the next
step. Certainly, Yang should tell Alice the photons operated with
Hadamard operations and they analyze their error rate after the
other agents publish their operations on these photons.

(10') Alice inserts randomly the decoy photons in the sequence
$S_C$ and then sends it to the last agent Zach. They analyze the
security of this transmission with the decoy photons. If they
conclude that the quantum line is secure, they continue their
communication to next step; otherwise, they discard the results
and repeat the quantum communication from the beginning.

(11') Alice selects a subset of photons as the samples for
eavesdropping check and chooses one of the four unitary operations
randomly on each sample. For other photons in the $S_A$ sequence
(except for those for eavesdropping check), Alice encodes her
message $M_A$ on them with the four unitary operations, and then
sends the sequence $S_A$ to Zach.

(12') Zach performs the Bell-state measurements on the EPR photon
pairs and reads  out the combination of the operations done by
Alice and Bob, i.e., $U_Z = U_A \otimes U_B \otimes U_C \otimes
\cdots \otimes U_Y$.

(13') Alice and Zach finish the error rate of the samples selected
by Alice, and determine whether there are eavesdroppers monitoring
the quantum channel when the $S_A$ is transmitted from Alice to
Charlie. For this end, Alice first requires all the agents to
publish the operations done on the correlated photons in the
sequence $S_C$, and then she requires Zach to tell her the results
of the Bell measurements. If the transmission of the $S_A$
sequence is secure, Alice tells all the agents to collaborate for
reading out the message $M_A$, otherwise they will abandon the
results of the transmission.

The security of this multiparty quantum secret splitting protocol
is same as that with two agents as each process for the
transmission of the photons is same as that in BBM92 QKD protocol
\cite{BBM92}. That is, the process for secure transmission of
photons in BBM92 QKD is repeated $N+2$ times. So this  multiparty
quantum secret splitting is secure also.

\section{MQSTS of an arbitrary M-particle state with  EPR pairs}

The state of an M-particle quantum system $x$ can be written as
following:
\begin{eqnarray}
\vert \psi\rangle_x=\sum_{i,j,\ldots, k\in \{0,1\}} C_{ij\ldots
k}\vert \underbrace{ij\ldots k}_M\rangle,
\end{eqnarray}
where
\begin{eqnarray}
\sum_{i,j,\ldots, k\in \{0,1\}} \vert C_{ij\ldots k}\vert ^2=1.
\end{eqnarray}
The state $\vert \psi\rangle_x$ can be teleported with M EPR pairs
completely \cite{Yangcp}. That is, the protocol for quantum secret
splitting of classical message discussed above can be modified to
accomplish the task of multiparty quantum state sharing (MQSTS)
with quantum teleportation. We depict it with three partes (Alice,
Bob and Charlie) in brief as follows.

\begin{figure}[!h]
\begin{center}
\includegraphics[width=4cm,angle=0]{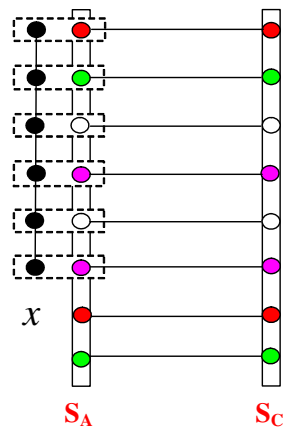} \label{f1}
\caption{ The Bell measurements performed by Alice. The bold lines
connect the two photons in an EPR pair or the M-particle quantum
system $x$. The panes with dashed lines are the Bell-state
measurements.}
\end{center}
\end{figure}

(I) Alice and Bob share an ordered $N$ EPR pairs privately using
the same way as  that for quantum secret splitting discussed
above.

(II) Bob chooses randomly one of the four unitary operation $U_i$
($i=0,1,2,3$) to encrypt each photon in the $S_C$ sequence, and
then sends the sequence to Alice. She picks up $k+j$ EPR pairs and
requires Bob to tell her the operations on these samples. She
analyzes the security of the transmission for the sequence $S_C$
by measuring $K$ pairs with MB Z or X on the two photons in each
pair. Then she performs a single-photon measurement on one photon
in each of the other $j$ EPR pairs in the samples. She uses the
photons collapsed as the decoy photons for checking eavesdropping
in the next step by inserting them in the sequence $S_c$. She
sends the sequence $S_C$ to Charlie

(III) Alice and Charlie finish the eavesdropping check for the
transmission of the $S_C$ sequence with the decoy photons. If the
error rate is low, they continue to the next step, otherwise they
abandon the results transmitted and repeat the quantum
communication from the beginning.

(IV) Alice picks out $M$ EPR pairs from the N ordered EPR photon
pairs ($N\geq M$), i.e., she select $M$ photon in the $S_A$
sequence in order. Then she performs a Bell-state measurement on
each photon and a particle from the quantum system $x$, shown in
Fig.1.

(V) Alice publishes the results of the Bell-state measurements in
public, and Charlie can reconstructs the quantum system $x$ with
the help of Bob's.

\begin{table}
\label{table1}
\begin{center}
\caption{The relation between the local unitary operations and the
results $\vert \phi\rangle_{X_1A_1}$ and $\vert
\phi\rangle_{X_2A_2}$. $\vert \Phi\rangle_{C_1C_2}$ is the state
of the two particles hold in the hand of Charlie after all the
measurements are done by Alice; $U_i\otimes U_j$ are the local
unitary operations with which Charlie can reconstruct the unknown
arbitrary entangled state $\vert \Phi\rangle_{X_1X_2}$. }
\begin{tabular}{cccc|ccc|cc}\hline
 $\vert \phi\rangle_{X_1A_1}$  & & $\vert \phi\rangle_{X_2A_2}$& & & $(U_{B_1}^{-1} \otimes U_{B_2}^{-1}) \vert \Phi\rangle_{C_1C_2}$ & & &
$U_i\otimes U_j$\\\hline 
 $\phi^+$  & & $\phi^+$ & & & $a\vert 11\rangle - b\vert 10\rangle - c\vert 01\rangle + d\vert
00\rangle$ & & & $U_3\otimes U_3 $
 \\ 
 $\phi^+$  & & $\phi^-$ & & & $a\vert 11\rangle + b\vert 10\rangle - c\vert 01\rangle - d\vert
00\rangle$ & & & $U_3\otimes U_2 $
\\ 
 $\phi^+$  & & $\psi^+$ & & & $a\vert 10\rangle - b\vert 11\rangle - c\vert 00\rangle + d\vert
01\rangle$ & & & $U_3\otimes U_1 $
\\ 
 $\phi^+$  & & $\psi^-$ & & & $a\vert 10\rangle + b\vert 11\rangle - c\vert 00\rangle - d\vert
01\rangle$ & & & $U_3\otimes U_0 $
\\ 
 $\phi^-$  & & $\phi^+$ & & & $a\vert 11\rangle - b\vert 10\rangle + c\vert 01\rangle - d\vert
00\rangle$ & & & $U_2\otimes U_3 $
\\ 
 $\phi^-$  & & $\phi^-$ & & & $a\vert 11\rangle + b\vert 10\rangle + c\vert 01\rangle + d\vert
00\rangle$ & & & $U_2\otimes U_2 $
\\ 
 $\phi^-$  & & $\psi^+$ & & & $a\vert 10\rangle - b\vert 11\rangle + c\vert 00\rangle - d\vert
01\rangle$ & & & $U_2\otimes U_1 $
\\ 
 $\phi^-$  & & $\psi^-$ & & & $a\vert 10\rangle + b\vert 11\rangle + c\vert 00\rangle + d\vert
01\rangle$ & & & $U_2\otimes U_0 $
\\ 
 $\psi^+$  & & $\phi^+$ & & & $a\vert 01\rangle - b\vert
00\rangle - c\vert 11\rangle + d\vert 10\rangle$ & & & $U_1\otimes
U_3 $
\\ 
 $\psi^+$  & & $\phi^-$ & & &  $a\vert 01\rangle + b\vert 00\rangle - c\vert 11\rangle - d\vert
10\rangle$ & & & $U_1\otimes U_2 $
\\ 
 $\psi^+$  & & $\psi^+$ & & & $a\vert 00\rangle - b\vert
01\rangle - c\vert 10\rangle + d\vert 11\rangle$ & & & $U_1\otimes
U_1 $
\\ 
 $\psi^+$  & & $\psi^-$ & & & $a\vert 00\rangle + b\vert 01\rangle - c\vert 10\rangle - d\vert 11\rangle$ & & & $U_1\otimes
U_0 $
\\ 
 $\psi^-$  & & $\phi^+$ & & &  $a\vert 01\rangle - b\vert 00\rangle + c\vert 11\rangle - d\vert
10\rangle$ & & & $U_0\otimes U_3 $
\\ 
 $\psi^-$  & & $\phi^-$ & & & $a\vert 01\rangle + b\vert 00\rangle + c\vert 11\rangle + d\vert
10\rangle$ & & & $U_0\otimes U_2 $
\\ 
 $\psi^-$  & & $\psi^+$ & & &  $a\vert 00\rangle - b\vert 01\rangle + c\vert 10\rangle - d\vert
11\rangle$ & & & $U_0\otimes U_1 $
\\ 
 $\psi^-$  & & $\psi^-$ & & & $a\vert 00\rangle + b\vert 01\rangle + c\vert 10\rangle + d\vert
11\rangle$ & & & $U_0\otimes U_0 $
\\\hline
\end{tabular}
\end{center}
\end{table}

In fact, this quantum state sharing protocol is just a process
that Alice  teleports the M-particle quantum system to Charlie who
does not know the states of the quantum channel, i.e., those of
the EPR pairs shared. Let us use a two-particle quantum system as
an example to describe the principle in detail. The other case is
the same as it with or without a little modification.

Assume that the unknown two-particle state $\vert \psi\rangle_x$
is $\vert \psi\rangle_x \equiv \vert \phi\rangle_{X_1X_2}=a\vert
00 \rangle +b\vert 01 \rangle + c\vert 10 \rangle + d\vert
11\rangle$, here $\vert a\vert^2 + \vert b\vert^2 + \vert c\vert^2
+ \vert d\vert^2 =1$. Alice and Charlie first pick out two EPR
pairs from the $N$ ordered EPR pairs shared, say $\vert
\phi\rangle_{A_1C_1}$ and $\vert \phi\rangle_{A_2C_2}$, and then
Alice performs the Bell-state measurements on the particles $X_1$
and $A_1$, and  $X_2$ and $A_2$, respectively. She publishes the
outcomes. If Charlie knows the states of the EPR pairs shared
between him and Alice, he can recover the unknown quantum state
$\vert \psi\rangle_x$ with the local unitary operations $U_i
\otimes U_j$. Here the operations $U_i$ and $U_j$ are the two
local unitary operations done by Charlie on the two photons $C_1$
and $C_2$, respectively. As the original states of the EPR pairs
are $ \vert \psi^- \rangle _{AC} =\frac{ 1}{\sqrt{2}}(\vert
0\rangle _{A}\vert 1\rangle _{C} - \vert 1\rangle _{A}\vert
0\rangle _{C})$, Charlie should obtain the operations $U_B$ done
by Bob on the photons in the $S_c$ sequence. That is, Charlie can
recover the unknown quantum state only when he collaborate with
Bob. The relation between the local unitary operations and the
information published by Alice is shown in Table I. Here $U_{B_1}$
and $U_{B_2}$ are the two local unitary operations done by Bob on
the EPR pairs $\vert \phi\rangle_{A_1C_1}$ and $\vert
\phi\rangle_{A_1C_1}$, respectively. $U_{B_1}^{-1}\otimes
U_{B_1}=I$ and $U_{B_2}^{-1}\otimes U_{B_2}=I$. $(U_{B_1}^{-1}
\otimes U_{B_2}^{-1}) \vert \Phi\rangle_{C_1C_2}$ is the state of
the two photons $C_1$ and $C_2$ after the operations
$U_{B_1}^{-1}$ and $U_{B_2}^{-1}$ on them respectively.

The way for generalize this MQSTS protocol to the case with $N$
agents is similar to that in the multiparty quantum secret
splitting.

\section{Discussion and summary}

As discussed in Refs. \cite{QSDCNetwork,two-step,Wangc,QOTP}, the
direct transmission of secret message should resort to the
transmission of quantum data block. The parties should confirm the
security of the quantum channel before the message is encoded on
the states. In present multiparty quantum secret splitting and
quantum state sharing protocols, the states are transmitted in
quantum data block. The good feature is that the error correction
\cite{book} and the quantum privacy amplification \cite{QAP} can
be done on the states directly, which will ensure the security of
those quantum communication protocols in a practical channel. In a
loss channel, Alice and Charlie should prevent Bob from
eavesdropping with quantum teleportation, same as the two step
QSDC protocol \cite{two-step}. That is, Charlie will determine
which position in the $S_C$ sequence has no photon and tell Alice
not perform encoding the message on the correlated photon in the
$S_A$ sequence.

In present multiparty quantum secret splitting protocol, each EPR
pair can carry two bits of information and its security is ensured
by the eavesdropping checks with the single-photon measurements
between the sender and the agents. The efficiency for qubits
$\eta_q\equiv \frac{q_u}{q_t}$ approaches the maximal value 100\%
as almost all the EPR pairs are useful for carrying the message in
principle, and here $q_u$ is the useful qubits and $q_t$ is the
total qubits transmitted. The total efficiency $\eta_t$ in this
protocol also approaches 100\% as the classical information
exchanged is not necessary except for the eavesdropping checks.
$\eta$ is defined as \cite{longqkd,Cabello}
\begin{eqnarray}
\eta_t=\frac{b_m}{q_t+b_t},
\end{eqnarray}
where $b_m$ is the number of bits in the message $M_A$, $q_t$ and
$b_t$ are the qubits used and the classical bits exchanged between
the parties in the quantum communication, respectively.

For sharing the unknown quantum state, the agent Bob plays a role
for preparing the quantum source and encrypting a classical
information on them also. The last agent will recover the unknown
quantum state with the help of the other agents as he keeps the
only quantum system for the reconstruction after all the Bell
measurements are done by Alice. In this way, the present
multiparty quantum state sharing protocol is asymmetrical, which
is same as that in Ref. \cite{zhanglm} and is different to the
symmetrical protocols \cite{dengpra,dengmQSTS} in which any one in
the agents can act as the receiver with the help of others'.
Certainly, any cheating done by the agents including the receiver
can be detected if the sender and the agents accomplish a honesty
check before the agents cooperate to recover the unknown states.
That is, Alice inserts randomly some samples in the unknown
quantum system $x$ and requires the agents to recover their states
and measure them before they obtain the quantum information, same
as that in Ref. \cite{deng2005}. The total efficiency $\eta_t$ of
this multiparty quantum state sharing protocol is about 50\% as
Alice should publish two bits of classical information for two
bits of quantum information.

In summary, we present a way for multiparty quantum secret
splitting with an ordered $N$ EPR pairs following some ideas in
Ref. \cite{two-step,Wangc,zhangPLA}. It is secure and each of the
EPR pairs can carry two bits of message. Its efficiency for qubits
and the total efficiency both approach the maximal value 100\%.
Moreover, we modify it for multiparty quantum state sharing
(MQSTS) of an arbitrary $M$-particle entangled state based on
quantum teleportation. For the latter, the task is completed with
the quantum channel of two-particle entanglement  and the
two-particle Bell-state measurements, which make it more
convenient in a practical application than others as it is
difficult for producing and measuring a multipartite entangled
states at present
\cite{Mentanglement1,Mentanglement2,Mentanglement3}.

\section{Note}
This paper appeared in PLA. However, there is a security loophole
in the original manuscript (also for other schemes with the
similar principle). It is necessary for us to forbid the dishonest
agent eavesdrop the quantum communication with a fake signal (a
Bell state) and cheat. In this way, the boss Alice should have the
capability of detecting cheat. In this revision, we added some
procedures for detecting cheat and made this protocol be secure.

\section{ACKNOWLEDGEMENTS}
This work is supported by the National Natural Science Foundation
of China under Grant Nos. 10447106, 10435020, 10254002, A0325401
and 10374010, and Beijing Education Committee under Grant No.
XK100270454.

\end{document}